\documentstyle[12pt]{article}
\newcommand{\ltsim}{\displaystyle\mathop{<}_{\sim}}

\begin{document}

\begin{flushright}  
\setlength{\baselineskip}{2.6ex}
TRI--PP--96--1\\
Feb 1996\\
Revised May 1996
\end{flushright}
\vspace{0.8cm}

\begin{center}
{\huge  \bf Light hadron masses with a \\ tadpole-improved \\
next-nearest-neighbour lattice \\ fermion action}\\
\vspace{0.8cm}
{H.R. Fiebig}\\
\vspace{0.3cm}
{\small\em Physics Department, FIU - University Park, Miami, Florida 33199 USA}\\
\vspace{0.8cm}
{R.M. Woloshyn}\\ 
\vspace{0.3cm}
{\small\em TRIUMF, 4004 Wesbrook Mall, Vancouver, B.C. Canada, V6T 2A3}
\end{center}

\vspace{1.0cm} \begin{center} {\bf Abstract} \end{center}

\setlength{\baselineskip}{3ex}

\noindent
Calculations of hadron masses are done in quenched approximation using
gauge field and fermion actions which are both corrected for discretization
errors to $O(a^2)$ at the classical level and which contain tadpole
improvement factors. The fermion action has both nearest-neighbour and 
next-nearest-neighbour couplings in the kinetic and Wilson terms.  
Simulations done at lattice spacings of $0.27$ and $0.4$fm yield hadron masses
which are already quite close to experimental values.
The results are compared to Wilson action calculations done at comparable
lattice spacings.

\setlength{\baselineskip}{5ex}

\newpage

\section{Introduction}


During the past several years there has been renewed interest in the use of
improved lattice actions. Many calculations have been done using the so-called
clover action\cite{mart,ukqcd,ape} motivated by the pioneering work of 
Sheikholeslami and Wohlert\cite{sw}.
Recently the move toward improved actions has been given even more impetus
by the work of Lepage and co-workers\cite{alfor_coars,alfor_d234}
which suggests that with tadpole improvement\cite{lep_mac}, 
calculations can be done quite accurately even on rather
coarse lattices. In this note we report on calculations done with a simple
tadpole-improved next-nearest-neighbour fermion action which support this
suggestion.

The essential idea of improved actions is that by including terms that are
nonleading (in powers of lattice spacing) one can reduce discretization 
errors. Of course, the choice of action is not unique. The approach of
Sheikholeslami and Wohlert\cite{sw} is to impose the minimal 
on-shell improvement condition\cite{lw} and they showed that $O(a)$ errors 
could be removed from physical
observables by the use of the so-called clover action. An advantage of this
action is that to $O(a)$ the familiar Wilson plaquette action may be used for
the gauge field. However, as Lepage and co-workers\cite{alfor_coars,alfor_d234}
have shown, a significant
gain can be made in improving the gauge field actions by incorporating tadpole
factors\cite{lep_mac} in the weighting 
coefficients of the nonleading terms. This, for
example, leads to restoration of rotational invariance\cite{alfor_coars} of the
static potential even at lattice spacings of order $0.4$fm. In 
ref.\cite{alfor_d234}
Alford et al extended the tadpole improvement program to the light
quark sector, introducing the $D234$ fermion action which, 
at the classical level,
is corrected to $O(a^2)$. A feature of the $D234$ action is that it contains
both clover and next-nearest-neighbour terms in addition to the terms
appearing in the Wilson action. 

In this work we consider the use of a
simple alternative to the $D234$ action which dispenses with the clover
term altogether. This is the next-nearest-neighbour fermion
action\footnote{The use of this action with tadpole improvement
has also been considered independently by
Lee and Leinweber\protect\cite{LeeLeinw}.}
in which both kinetic and Wilson terms have been corrected
at tree-level to $O(a^2)$\cite{nnn}.
In addition, tadpole factors are included in the next-nearest-neighbour
terms. These work to remove, in a mean field sense, discretization errors
due to tadpole-like couplings induced by the lattice description of the
gauge field\cite{lep_mac}. In conjunction with this fermion action we use a 
gauge field action that has been analogously improved, 
that is, $O(a^2)$ tree-level improvement plus tadpole factors.

Here we report some results for light-quark (u, d and s) meson and baryon
masses calculated on lattices with lattice spacings of $0.4$ and $0.27$fm.
In general, our results are fairly close
to experimental values and are compatible with Wilson action 
calculations (see, for example, Ref.\cite{gf11})
done at lattice spacings of about $0.1$fm.

For purposes of comparison simulations have also been carried out 
with the Wilson action at lattice spacings matched to those of the 
improved action calculations. As expected, on our coarse lattices,
the improved action results are much closer to those obtained 
with the Wilson action at small lattice spacing.
What we observe is that the major difference between the improved action
and the Wilson action is in the relative scale between the meson and baryon
sectors. Mass  ratios within the  meson sector and within the baryon
sector depend much less on the choice of action. 
 
\section{Method}

%
%
The $SU(3)$ gauge fields are described by an action that contains both
4-link square plaquettes (pl) and planar 6-link rectangular plaquettes (rt). As
shown in \cite{weisz} the 6-link rectangles are sufficient to remove $O(a^2)$
errors at the classical level. In addition tadpole factors are introduced
into the weighting of the 6-link term. The action is
\begin{equation}
S_G(U) = \beta\left[\sum_{pl}(1-\frac13\mbox{Re}\mbox{Tr}U_{pl})
             +C_{rt}\sum_{rt}(1-\frac13\mbox{Re}\mbox{Tr}U_{rt})\right]
\label{eq1}\end{equation}
where $U_{pl}$ are the square plaquettes and $U_{rt}$ are the planar
6-link plaquettes. The coefficient $C_{rt} = -1/20U_0^2$ includes the tadpole
factor
\begin{equation}
U_0 = \langle\frac13\mbox{Re}\mbox{Tr}U_{pl}\rangle^{1/4} \,.
\label{eq2}\end{equation}
The first term of (\ref{eq1}) is just the Wilson action.

For the fermions, the Wilson action augmented by next-nearest-neighbour
couplings\cite{nnn} in both the kinetic and Wilson terms is used. Including 
tadpole factors the action is
\begin{eqnarray}
S_F(\bar{\psi},\psi;U) &=& \sum_{x,\mu}\frac{4}{3}\kappa\left[
\bar{\psi}(x)(1-\gamma_{\mu})U_{\mu}(x)\psi(x+\mu)\right. \nonumber \\
 & & \left. +\bar{\psi}(x+\mu)(1+\gamma_{\mu})
U_{\mu}^{\dagger}(x)\psi(x)\right]\nonumber \\
 & & -\sum_{x,\mu}\frac{1}{6}\frac{\kappa}{U_0}
\left[\bar{\psi}(x)(2-\gamma_{\mu})
U_{\mu}(x)U_{\mu}(x+\mu)\psi(x+2\mu)\right. \nonumber \\
 & & \left.+\bar{\psi}(x+2\mu)(2+\gamma_{\mu})
U_{\mu}^{\dagger}(x+\mu)U_{\mu}^{\dagger}(x)\psi(x)\right] \nonumber \\
 & & -\sum_{x}\bar{\psi}(x)\psi(x).
\label{eq3}\end{eqnarray}
With the coefficients as in (\ref{eq3}), $\kappa_{critical} = 1/8$
at the tree level,
the same as for the Wilson fermion action. The Wilson action is recovered
by replacing the coefficient $4/3$ by $1$ and dropping next-nearest-neighbour
terms.

A feature of next-nearest-neighbour action (3), which it shares with the 
$D234$ action, is the presence of unphysical states in the free quark
propagator with a massless dispersion relation very similar to that
given by Alford et al\cite{alfor_d234}. One might wonder about the 
effect of such unphysical singularities.
In fact, the near identity in the results reported by 
Alford et al\cite{alfor_d234} and by Collins et al\cite{collins} who use
a clover action which has no doublers suggests to us that the singularity
structure of the tree-level
propagator may not be very crucial in determining the ability of an action
to describe hadron masses. A posteriori
the hadron masses which we calculate show
no obvious effect that can be linked to
the unphysical states of the free propagator although this is something that
merits further study. 
                           
%
%
Calculations were carried out in quenched approximation at two different
values of $\beta$, $6.25$ and $6.8$ for the improved action and $4.5$
and $5.5$ for the Wilson action. These values were chosen so that
lattice spacings determined from the string tension would match for the 
two actions\cite{hdt}.
These lattice spacings are $0.4$fm and $0.27$fm for the smaller
and larger $\beta$-values respectively. The lattice sizes used were $6^3\times
12$ and $8^3\times14$.

Gauge field updating was done using the Cabbibo-Marinari pseudo-heat\-bath.
Periodic boundary conditions were used for the gauge field in all directions.
The lattice was thermalized for $4000$ sweeps then configurations were used
every $250$ sweeps in the case of the improved action and every $200$ sweeps
for the Wilson action.

Quark propagators were calculated for a range of $\kappa$ values in each
simulation. A stabilized biconjugate gradient algorithm\cite{bicon} 
was used for these
calculations. Periodic boundary conditions were imposed on the quark fields
in spatial directions but in the time direction a Dirichlet or fixed 
boundary condition was used. This allows mass measurements to be made further
from the source than with periodic boundary conditions. This is an important
consideration given the relatively small number of time slices. The source
position was fixed to be two time steps in from the boundary in all 
simulations.

Meson and baryon correlators were calculated using standard local 
interpolating fields.
\begin{equation}
\chi^{(\Gamma)}(x) = \bar{\psi}(x)\Gamma\psi(x)
\label{eq4a}\end{equation}
with $\Gamma=\gamma_5,\gamma_{\mu}$ for the mesons,
\begin{equation}
\chi^{(N)}_{i}(x) = \epsilon_{abc}{\psi}_{a,i}(x)
\left[\psi_{b}^T(x)C\gamma_5\psi_{c}(x)\right]
\label{eq4b}\end{equation}
for the nucleon, where $C$ is the charge conjugation matrix, and
\begin{equation}
\chi^{(\Delta)}_{ijk}(x) = \epsilon_{abc}
\psi_{a,i}(x)\psi_{b,j}(x)\psi_{c,k}(x)
\label{eq4c}\end{equation}
for the isobar.

As is well known, smeared operators can be used to enhance the overlap of
the interpolating field with the ground state. This allows ground state masses
to be extracted closer to the source point where statistical fluctuations
are less severe. In fact, a smeared sink, although maybe less effective
than a smeared source, can be implemented at very little cost. Therefore,
correlators were constructed for both local and smeared sinks with
local sources. Gaussian smearing\cite{smear} was used. The smearing function is
\begin{equation}
\Phi = (1+\alpha H)^n
\label{eq5a}\end{equation}
where
\begin{equation}
H(\vec{x},\vec{y};t) = \sum_{i=1}^{3}
\left[U_{i}(\vec{x},t)\delta_{\vec{x},\vec{y}-\hat{i}}
+ U_{i}^{\dagger}(\vec{x}-\hat{i},t)\delta_{\vec{x},\vec{y}+\hat{i}}\right]\,.
\label{eq5b}\end{equation}
The smearing parameters were fixed at $n = 6$ and $\alpha = 2$
for all simulations. No attempt was made to optimize the smearing parameters 
for this exploratory calculation.

The lattice details for the calculations are summarized in Table~\ref{tab1}.

\section{Results and Conclusion}

Masses were calculated using an analysis procedure motivated by
Bhattacharya et al\cite{lanl}. 
For each channel the correlation functions $G(t)$
were configuration averaged and the effective mass function $M_{\mbox{eff}}(t) =
\ln(G(t)/G(t+1))$ was calculated. Then a combined effective mass function
was computed by a weighted average of the effective mass functions obtained
from local-local and local-smeared correlators. Since the local-local 
correlator overestimates the gound state mass and the local-smeared correlator
underestimates it, this average helps to enhance the plateau of the
effective mass. The mass is then determined by averaging the combined mass
function over some time interval. Except for the baryon spin-$3/2$ channel,
it was found that compatible masses could be obtained using time averages
starting 2 or 3 time steps away from the source.

For each simulation the masses were extrapolated as a function of pion
mass to the chiral limit $M_{\pi} = 0$. The choice of extrapolation
function was either
\begin{equation}
M = M_0+cM_{\pi}^2
\label{eq6a}\end{equation}
or
\begin{equation}
M = M_0+cM_{\pi}^2+dM_{\pi}^3 \,.
\label{eq6b}\end{equation}
The criterion was that the cubic form (which is motivated by chiral
perturbation theory\cite{labren}) was used whenever the coefficient d could be 
determined to be nonzero within the statistical errors. If the mass data
showed no evidence of a cubic term then the quadratic form was used.

The errors in masses and in mass ratios were estimated using a bootstrap
procedure. For each simulation $500$ bootstrap samples were chosen from the
original sample and analyzed for masses and mass ratios. The quoted errors
on observables are one half the difference between the $16^{\mbox{th}}$
and $84^{\mbox{th}}$
percentile values found in the bootstrap distribution for that observable.

In addition to the u,d sector we are also interested in the strange quark
sector. To fix $\kappa_s$, (i.e., the strange quark mass) the condition
$K^*/K$ equals the experimentally observed value $1.8$ was used. Generally
speaking $\kappa_s$ does not coincide with one of our chosen $\kappa$
values. This requires an interpolation (or extrapolation) which was done
linearly in $\kappa$ using the  two $\kappa$ values nearest $\kappa_s$. 

As a representative sample of our results we show the $\rho$-meson, nucleon and
delta masses as a function of $M_{\pi}^2$ in Fig.~1. Also shown are the
extrapolations to the chiral limit. Mass ratios extrapolated to the chiral
limit are given in Table~\ref{tab2}. Meson masses are given with respect to the
$\rho$-meson mass and baryon masses with respect to the nucleon mass. The
ratio $M_N/M_{\rho}$ then sets an overall scale of baryon masses relative
to meson masses and this seems to be the quantity most effected by
discretization errors.
By $0.27$fm the improved action results are fairly close to
experiment and are compatible with Wilson action calculations done at small
lattice spacing\cite{gf11}.  

The quantity $J = M_V\,dM_V/{dM_P}^2$ at $M_V/M_P=1.8$
was introduced by Lacock and Michael\cite{lm} as a measure of the
relative quark mass dependence of pseudoscalar and vector mesons. Empirically
this value is very close to 0.5 reflecting the fact that $M_V^2 - M_P^2$ is
almost constant. Quenched lattice QCD simulations at small lattice spacings
tend to give values around 0.37 which was interpreted in \cite{lm} as a
failure of the quenched approximation. Our improved action results are
consistent with the previous determinations.

In the continuum limit of full lattice QCD it is expected that the lattice
spacing determined from all physical quantities will be the same. In a 
quenched calculation there is no reason why this
should also be true. Nonetheless it is still expected that there will 
still be a scaling region in which the
ratio of lattice spacings is constant. In Fig.~2 we have compiled some 
results for the ratio of lattice spacing extracted from the string tension
to the lattice spacing determined by the ${\rho}$-meson mass.
The improved action results include  our calculations as well as
the values reported by
Alford et al\cite{alfor_d234} for the $D234$ action and by Collins et 
al\cite{collins} using
a tadpole-improved clover fermion action with an $O(a^2)$ tadpole-improved
gauge field action.
The improved action and Wilson action results show the same qualitative
behaviour only shifted in lattice spacing by about a factor of 3.
Unfortunately it is only the last two points at the smallest value of
$M_{\rho}a$ (i.e., the largest $\beta$) which show a hint of scaling. 
If this really is the onset of scaling it would correlate very well with
the onset of the weak coupling region as shown, for example, by the 
behaviour of the average plaquette. A remarkable feature seen in Fig.~2
is that different improved fermion actions exhibit a high degree of 
universality even in the non-scaling region.

If simulations done with improved actions on coarse lattices are to be useful
the results should extrapolate smoothly to the continuum limit. Calculations
in the light hadron sector
using tadpole improved actions of the type used in this work are still too
scarce to be able to make definitive statements. However, the ratio of nucleon
to $\rho$-meson mass has been calculated a number of times. The
results of tadpole improved actions\cite{alfor_d234,collins} and a 
sample of Wilson action results\cite{gf11,lanl,qcdpax}
are presented in Fig.~3.
It is encouraging that our improved 
action values at $0.27$ and $0.4$fm are compatible 
with the Wilson action results at smaller lattice spacing.
However, it is somewhat disconcerting that in this case 
they do not agree with the 
values given by
Alford et al\cite{alfor_d234}and by Collins et 
al\cite{collins}.
At present, the results
of Ref.\cite{alfor_d234,collins} would suggest a continuum limit value
for $M_N/M_{\rho}$ different from the Wilson action value which is not a
very palatable conclusion. It is clear that the improved action calculations
have to be pushed to smaller lattice spacing to clarify this situation.


In this work we consider the use of a next-nearest-neighbour fermion action,
corrected for discretization errors to $O(a^2)$ at the classical level, 
for use with the tadpole 
improvement program of Lepage and co-workers
\cite{alfor_coars,alfor_d234,lep_mac}. An analogously
improved gauge field action is used. Quenched calculations done at lattice
spacings of
$0.27$ and $0.4$fm yield hadron masses in the light quark sector which are
comparable to those obtained with the Wilson action at much smaller
(a${\ltsim}0.1$fm) lattice spacing and 
which are quite close to experimental values.
Comparison with Wilson action calculations
shows very clearly the positive effect of improvement.

\section{Acknowledgements}

It is a pleasure to thank S.R. Beane, F.X. Lee and H.D. Trottier
for helpful discussions.
This work was supported in part by the Natural Sciences and 
Engineering Research Council of Canada, and by NSF grant PHY-9409195.

\newpage
 

\bibliographystyle{unsrt}      

\newpage
 
\begin{table}[htp]
\begin{tabular*}{138mm}{@{\extracolsep{\fill}}ccccccc} \hline
\rule[-3mm]{0mm}{8mm}
Action & Lattice & $N_U$ & $\beta$ & $a_{st}$ &
 $\kappa$ & $\kappa_s$ \\ \hline \\
Improved & $6^3\times 12$ & 160 & 6.25 & 0.4 fm&
 0.162, 0.165, 0.168, & \\ & & & & & 0.171, 0.174 & 0.166  \\ \\
Improved & $8^3\times 14$ &  60 & 6.8  & 0.27fm&
 0.148, 0.150, 0.152, & \\ & & & & & 0.154, 0.156, 0.158 & 0.1558 \\ \\
Wilson   & $6^3\times 12$ & 160 & 4.5  & 0.4 fm&
 0.189, 0.193, 0.197, & \\ & & & & & 0.201, 0.205, 0.209, & \\
 & & & & & 0.213 & 0.205 \\ \\
Wilson   & $8^3\times 14$ &  90 & 5.5  & 0.27 fm&
 0.164, 0.168, 0.172, & \\ & & & & & 0.176, 0.180 & 0.178 \\ \\ \hline
\end{tabular*}
\caption{Lattice Details. $N_U$ is the number of gauge configurations and
$a_{st}$ is the lattice spacing determined from the string tension\cite{hdt}.
$\kappa_s$ is the hopping parameter corresponding to the strange quark mass.}
\label{tab1}\end{table}
 
\begin{table}[htp]
\begin{tabular*}{138mm}{@{\extracolsep{\fill}}cccccc} \hline
\rule[-3mm]{0mm}{8mm}
   & \multicolumn{2}{c}{Improved} & \multicolumn{2}{c}{Wilson} & Exp. \\
   & $\beta=6.25$ & $\beta=6.8$ & $\beta=4.5$ & $\beta=5.5$ \\ \hline \\
$M_{\rho}a_{\rho}$ & 1.19(5) & 0.90(5)         & 0.90(2) & 0.71(3) \\ \\
$a_{\rho}^{-1}$ & 648(27)MeV & 855(45)MeV & 858(15)MeV & 1085(46)MeV \\ \\
$a_{st}/a_{\rho}$  & 1.31(5) & 1.17(7)    & 1.73(3) & 1.50(6) &      \\ \\
$J$                & 0.43(8) & 0.38(7)    & 0.31(2) & 0.32(7) &      \\ \\
$K/{\rho}$         & 0.65(2) & 0.65(4)    & 0.61(1) & 0.61(4) & 0.64 \\ \\
$K^{\ast}/{\rho}$  & 1.17(3) & 1.16(4)    & 1.10(1) & 1.13(4) & 1.16 \\ \\
$\phi/{\rho}$      & 1.31(4) & 1.30(6)    & 1.20(2) & 1.24(6) & 1.32 \\ \\
$N/{\rho}$         & 1.55(6) & 1.36(9)    & 2.05(5) & 1.73(14)& 1.22 \\ \\
$\Delta/N$         & 1.34(4) & 1.38(11)   & 1.07(2) & 1.24(10)& 1.31 \\ \\
$\Sigma/N$         & 1.15(2) & 1.20(4)    & 1.05(1) & 1.10(5) & 1.27 \\ \\
$\Xi/N$            & 1.23(3) & 1.32(4)    & 1.09(1) & 1.15(7) & 1.40 \\ \\
$\Lambda/N$        &         & 1.17(4)    &         & 1.08(5) & 1.19 \\ \\
$\Omega^-/N$       & 1.58(4) & 1.67(10)   & 1.19(2) & 1.38(10)& 1.78 \\  \\ \hline
\end{tabular*}
\caption{Results of the calculations extrapolated to the limit $M_{\pi}=0$.}
\label{tab2}\end{table}
 
\newpage

\begin{center}
{\bf Figure Captions}
\end{center}

\begin{enumerate}

\item Nonstrange hadron masses in lattice units versus pion mass squared
(squares = $\rho$-meson, circles = nucleon, triangles = Delta).
The lines are the extrapolations to the chiral limit.
(a) Improved action, $\beta = 6.25$,
(b) Improved action, $\beta = 6.8$,
(c) Wilson action, $\beta = 4.5$,
(d) Wilson action, $\beta = 5.5$.

\item The ratio of the lattice spacing determined by the string tension
to the lattice spacing determined by the $\rho$-meson mass
$a_{st}/a_{\rho}$ 
versus $M_{\rho}a$ for the Wilson (solid
symbols) and Improved actions (open symbols). For the Wilson action
points below $M_{\rho}a=0.6$, the string tension results of Bali
and Schilling\cite{bali} were used.

\item The ratio $M_N/M_{\rho}$ versus $M_{\rho}a$ for the Wilson (solid
symbols) and Improved actions (open symbols).  

\end{enumerate}

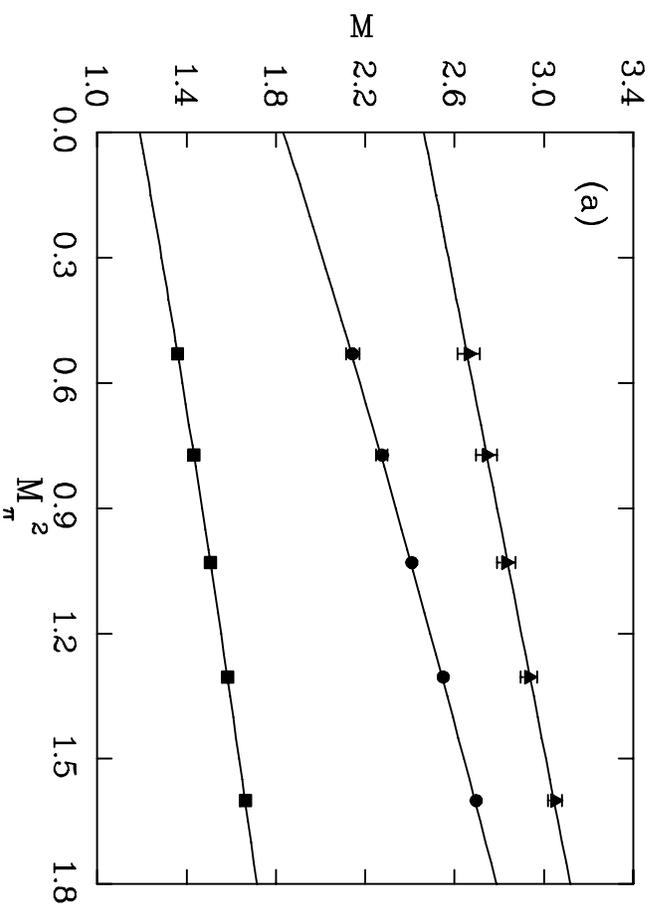
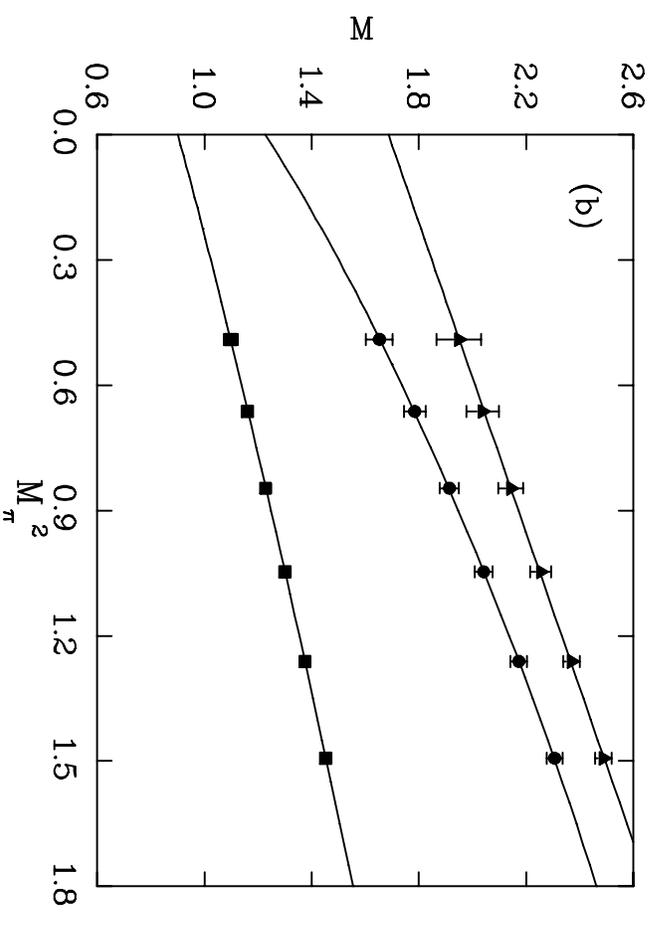
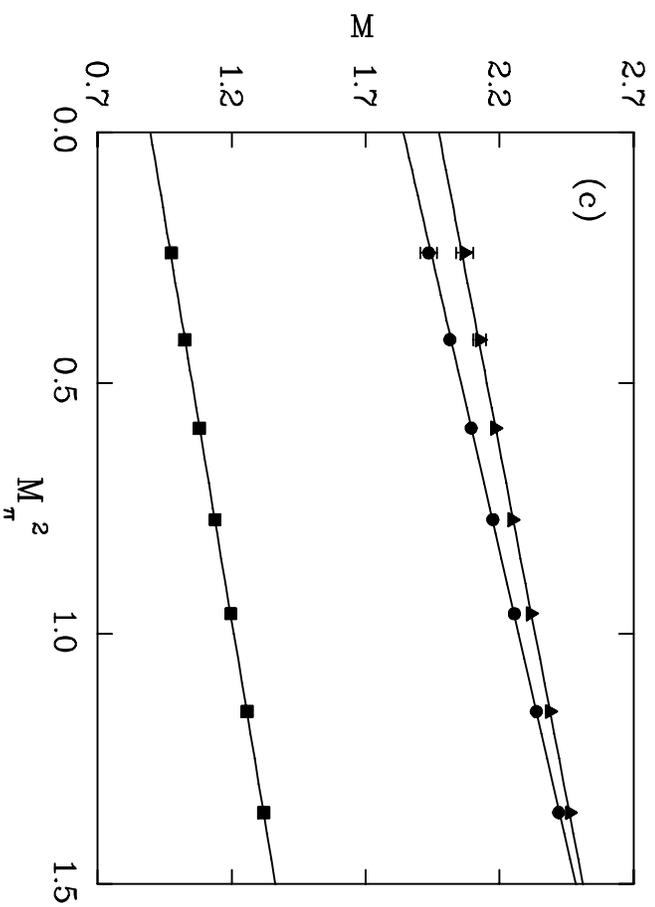
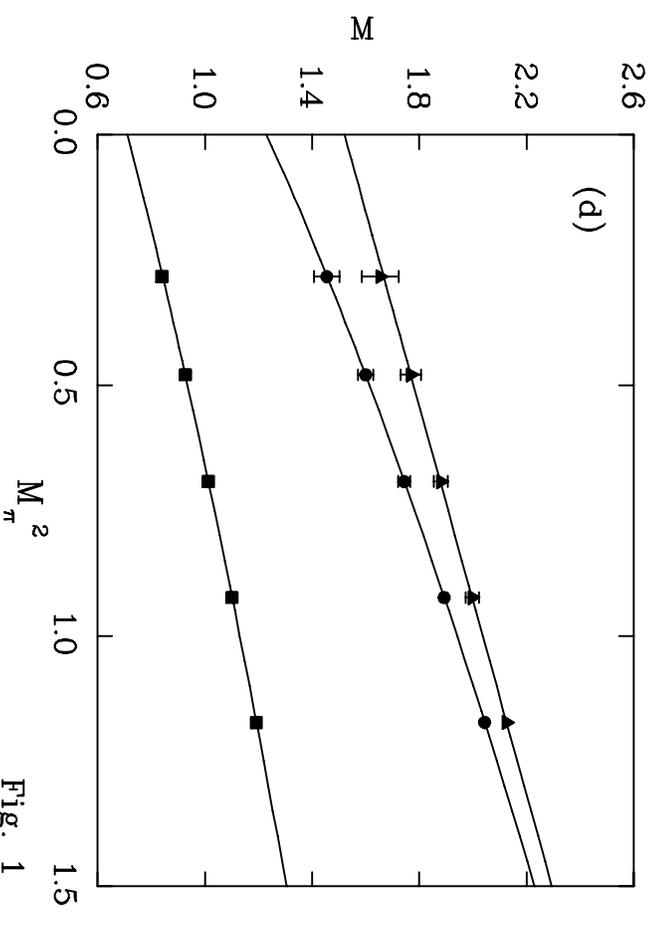
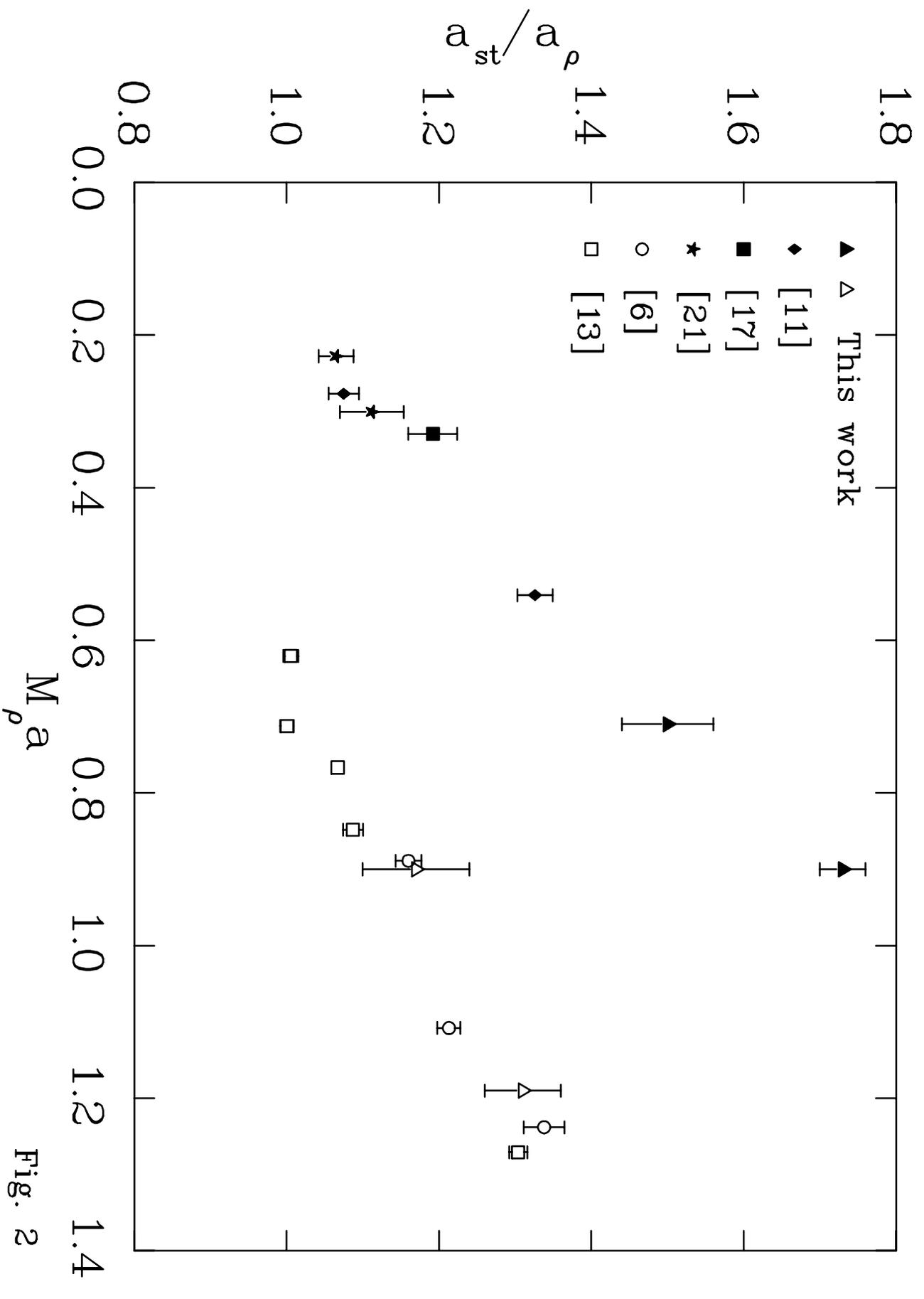
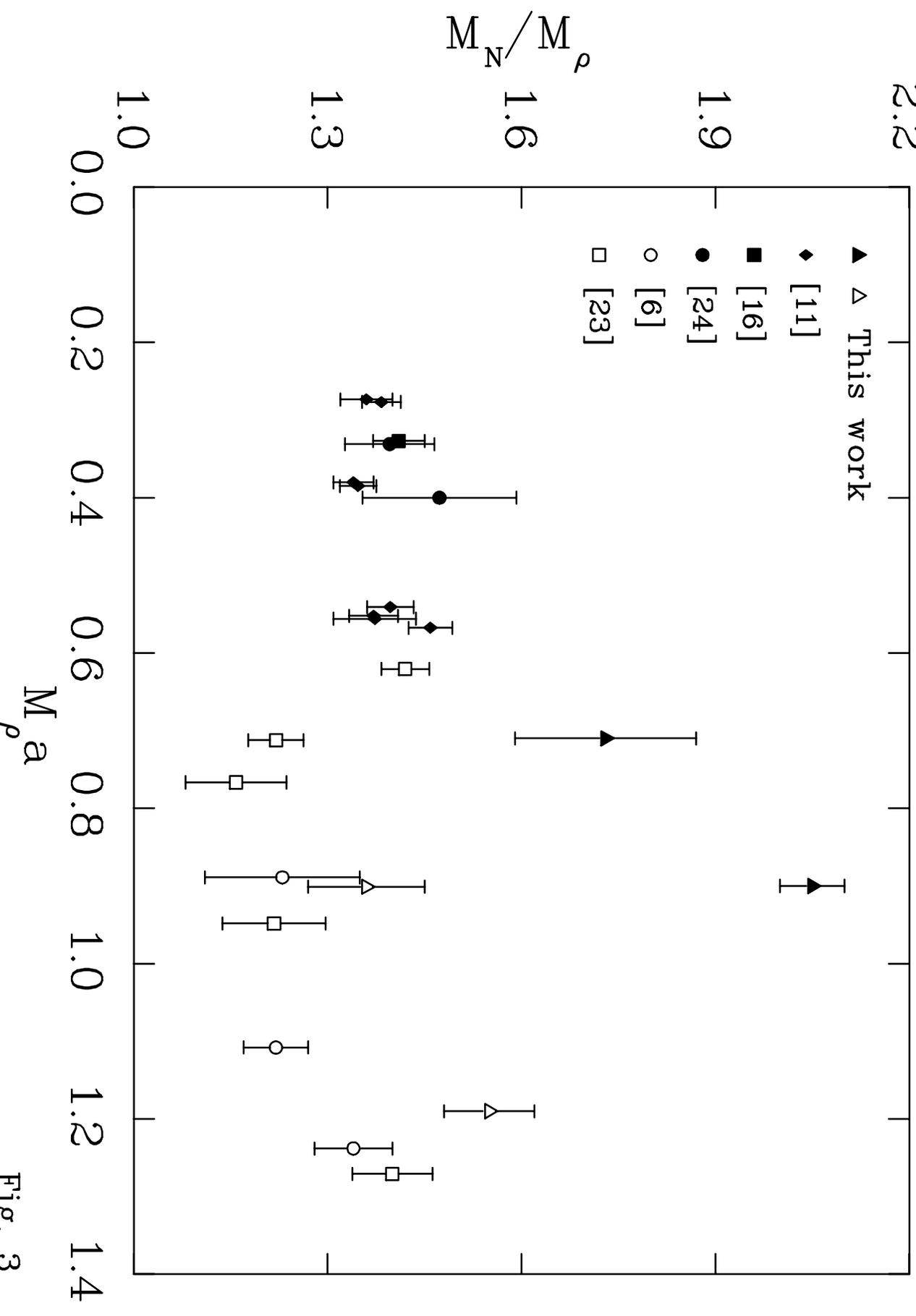
\end{document}